\documentclass{iopart}
\usepackage{psfig}
\begin{document}

\title{Numerical study of Schramm-Loewner Evolution in the random 3-state Potts model}

\author{C. Chatelain}
\address{
Groupe de Physique Statistique,\\
Institut Jean Lamour, UMR 7198,\\
Nancy-Universit\'e, CNRS,
BP~70239, Boulevard des aiguillettes,\\
F-54506 Vand{\oe}uvre l\`es Nancy Cedex, France}
\ead{chatelai@lpm.u-nancy.fr}

\begin{abstract}
We have numerically studied the properties of the interface induced in the
ferromagnetic random-bond three-state Potts model by symmetry-breaking boundary conditions.
The fractal dimension $d_f$ of the interface was determined. The corresponding
SLE parameter $\kappa$ was estimated to be $\kappa\simeq 3.18(6)$, compatible
with previous estimate. On the other hand, we estimated $\kappa$ independently
from the probability of passage of the interface at the left of a given point.
The numerical data are well reproduced by the Schramm theoretical prediction
and the fit leads to $\kappa\simeq 3.245(10)$, in agreement with the first estimate.
This provides evidences that the geometric properties of spin interfaces in
the random 3-state Potts model may be described by chordal ${\rm SLE}_\kappa$.
\end{abstract}

\pacs{05.10.Ln, 05.40.-a, 05.50.+q, 05.70.Jk}

\def\build#1_#2^#3{\mathrel{
\mathop{\kern 0pt#1}\limits_{#2}^{#3}}}

\section*{Introduction}
The assumption of conformal invariance in two-dimensional critical systems
has been very fruitful~\cite{Cardy96,DiFrancesco,Henkel99}.
Exact expressions for correlation functions, as well as values of critical exponents
for minimal models,
are among the most stringent predictions of conformal theory. Recently, the interest
has been focused on fractal curves that may be realized as interfaces induced
in a critical conformal-invariant system by imposing symmetry-breaking boundary
conditions~\cite{Cardy05,Bauer06,Lawler08}.
It is believed that the geometric properties of this kind of curves can be described
by a family of self-avoiding stochastic processes introduced by Schramm.
The self-avoidance property is obtained by iteratively removing, during the growth of
the curve, the hull formed from the curve itself and all non-reachable domains enclosed
by the curve. This removal is performed by a conformal transformation. As a consequence,
the construction relies on both the assumption of conformal invariance and on
Markovian property. The family of curves obtained in this way is parameterized by
a single parameter $\kappa$ corresponding to the intensity of the noise in the
underlying Brownian process.
\\

In a conformal field theory characterized by a central charge $c$, it has been predicted
that the interfaces induced by symmetry-breaking boundary conditions are described
by the chordal Schramm-L\"owner process of parameter $\kappa$ with
	\begin{equation}
	c={(3\kappa-8)(6-\kappa)\over 2\kappa}.
	\label{eq1}
	\end{equation}
The length $\ell$ of the interface is expected to scale anomalously with the lattice size
as $\ell\sim L^{d_f}$ where the fractal dimension is related to $\kappa$ by
	\begin{equation}
	d_f=1+{\kappa\over 8}.
	\label{Defdf}
	\end{equation}
The probability that a given point be on the left of the interface is given
by the Schramm formula:
	\begin{equation}
	\wp_{\rm left}(x,y)
	={1\over 2}-{\Gamma(4/\kappa)\over\sqrt\pi\Gamma((8-\kappa)/2\kappa)}
	\left({x\over y}\right) {}_2F_1\left({1\over 2},{4\over\kappa};
	 {3\over 2};-{x^2\over y^2}\right).
	\end{equation}
This formula is valid in the upper half-plane $y>0$ with fixed boundary conditions
on the real axis, changing for $x=0$. It can easily be extended to other
geometries by taking advantage of the conformal symmetry of the critical system
that allows to apply a conformal transformation from the upper-half plane
to another geometry. In the following, we will consider a critical system confined
in a square. The mapping of the point $z=x+iy$ in upper-half plane to the point
$\omega$ in the square is realized by the Schwartz-Christoffel conformal transformation
	\begin{equation}
	\omega = F(z,k) / K(k)\ \Leftrightarrow\ z={\rm sn}\ \! K(k)\omega
	\end{equation}
where $F(z,k)$ is the incomplete elliptic integral, $K(k)$ is the complete
elliptic integral and $k$ is related to the aspect ratio $s$ of the square
by
	\begin{equation}
	sK(k)=K'(k)
	\end{equation}
where $K'(k)$ is the associated elliptic integral.
\\

The critical properties of the $q$-state Potts model ($q\le 4$) defined by the
Hamiltonian~\cite{Potts}
	\begin{equation}
	-\beta{\cal H}=J\sum_{(i,j)}\delta_{\sigma_i,\sigma_j},\quad
	\sigma_i\in\{0,\ldots,q-1\}
	\end{equation}
are described by the conformal theory with central charge
	\begin{equation}
	c=1-{6\over m(m+1)}
	\end{equation}
where the integer $m$ is related to the number of states $q$ by the Nienhuis formula
	\begin{equation}
	q=4\cos^2\left({\pi\over m+1}\right).
	\end{equation}
One thus expects the geometric properties of the interfaces induced in the
Potts model to be described in the scaling limit by the chordal ${\rm SLE}_\kappa$
with the two possible values
	\begin{equation}
	\kappa_1={4\over 1-{1\over\pi}{\rm Arccos}\ \!(\sqrt q/2)},\quad
	\kappa_2=16/\kappa_1.
	\end{equation}
This prediction has been proved exactly by Smirnov in the case of
percolation~\cite{Smirnov01} ($q=1$) and latter for the Ising model~\cite{Smirnov06}
($q=2)$ in the Fortuin-Kasteleyn representation. In the case of the $q=3$ Potts model,
the fractal dimension of the interfaces between spin clusters has been shown to be
compatible within numerical accuracy with equation (\ref{Defdf}) and $\kappa_2=10/3$~\cite{Ziff}.
Predictions of ${\rm SLE}_\kappa$ have been tested more specifically for this model
by Gamsa and Cardy~\cite{Gamsa07}. These authors considered the interfaces between the
spin clusters and between the Fortuin-Kastelyn clusters and they extracted the parameter
$\kappa$ both from the scaling of the length of the interface and from the
Schramm formula for the left-passage probability. Several geometries and boundary conditions
were considered. As expected since $c=4/5$, Gamsa and Cardy measured the parameter
$\kappa_2=10/3$ for the spin interfaces and $\kappa_1=4.8$ for the interfaces
between the Fortuin-Kastelyn clusters, in agreement with the values expected for $c=4/5$
model. Unlike in the Ising model, certain ambiguities arise for the interfaces in the Potts model.
When fixing two different Potts states on the boundaries at the
left and at the right of the system, two interfaces are usually induced since a
cluster of the third state may appear between the left and right clusters.
Gamsa and Cardy generalized the Schramm formula for the left-passage probability 
to the case of two SLE curves~\cite{Gamsa05}. Unfortunately, in the 3-state Potts
model, the two curves sometimes collapse in a finite system and the latter formula
for the left-passage probability
does not hold. To circumvent this problem, Gamsa and Cardy have also studied
fluctuating boundary conditions: the left boundary is fixed to one of the Potts
states while at the right boundary, spins are free to take any of the two other
Potts states. With these boundary conditions, only one interface is induced and
the original Schramm formula for the left-passage probability can be applied.
Recently, the fractal dimension of the interface has been numerically shown
to be compatible with ${\rm SLE}_{\kappa}$ even for non-integer values of the number
of states $q$~\cite{Gliozzi}.
\\

The ferromagnetic random-bond 3-state Potts model is defined by the Hamiltonian
	\begin{equation}
	-\beta{\cal H}=\sum_{(i,j)} J_{ij}\delta_{\sigma_i,\sigma_j},\quad
	\sigma_i\in\{0,\ldots,q-1\},\quad J_{ij}>0
	\end{equation}
When the couplings are distributed according to the binary distribution
	\begin{equation}
	\wp(J_{ij})={1\over 2}\big[\delta(J_{ij}-J_1)+\delta(J_{ij}-J_2)\big]
	\end{equation}
the model is critical on the self-dual line
	\begin{equation}
	\big(e^{J_1}-1\big)\big(e^{J_2}-1\big)=q
	\end{equation}
Randomness is a relevant perturbation and thus alters the critical behavior.
In numerical simulations,
the ratio $r=J_1/J_2$ is usually fixed to the value that minimizes the scaling
corrections due to the cross-over with the pure and percolation fixed points.
The optimum $r^*$ can be determined as the value leading to the largest central
charge extracted from the scaling of the free energy density on the strip.
Even though conformal symmetry is broken for a single disorder realization, it
is restored after averaging over randomness. Magnetization profiles or correlation
functions in confined geometries have indeed been shown to transform covariantly
under conformal transformations~\cite{Berche}. Recently, the interface induced
by boundaries has been studied in the perspective of
${\rm SLE}_\kappa$~\cite{Jacobsen09}. The scaling of the interface length has
been determined both by transfer matrix techniques and Monte Carlo simulations.
From the fractal dimension $d_f$, Jacobsen {\sl et al.} estimated
$\kappa_2\simeq 3.208(24)$ for the spin interface. Duality, i.e.
$\kappa_1\kappa_2=16$, is satisfied within error bars. The results are compatible
with renormalization group arguments. Note that this estimate of $\kappa_2$
is not compatible with the value $\kappa\simeq 3.3370(4)$ given by relation
(\ref{eq1}) with the estimate of the central charge $c\simeq 0.8024(3)$.
\\

The Schramm formula has never been tested in the random Potts model. It has already
been considered in other random systems: Ising spin-glasses~\cite{Bernard07} and
the Solid-On-Solid model on a random substrate~\cite{Rieger09}. In the second
case, a large discrepancy was observed between the two estimates of $\kappa$ obtained
from the scaling of the interface length and from the left-passage probability.
Our aim is to check whether such a discrepancy exists for the random-bond Potts
model too. We will restrict ourselves to a square geometry with fluctuating boundary
conditions and to interfaces between spin clusters. In the first section, we present
the numerical details of the Monte Carlo simulations that we have performed. Then,
the fractal dimension of the interfaces is studied with an emphasis on the influence
of disorder on the value of $\kappa$. In the last section, we present our results for
the left-passage probability and the estimation of $\kappa$ that can be made from it.

\section{Details of the Monte Carlo simulations}
\label{sec1}
We have studied the random-bond 3-state Potts model using large-scale Monte Carlo
simulations. Square lattices were considered with fluctuating boundary conditions
and sizes $L=64$, 90, 128, 180, 256, 360 and 512.
Each Monte Carlo step consists into one iteration of the Metropolis algorithm and
one of the Swendsen-Wang algorithm~\cite{SWang}. 
This choice is motivated by the fact that the dynamics of an interface with the
Swendsen-Wang algorithm is very slow and may even be slower than with
local Monte Carlo algorithms (see for example Table~1 of \cite{Hasenbusch91}).
The combination of the two algorithms may help the system to escape
dynamical traps.
For equilibration of the system, between $25000$ ($L=64$) and $57810$ ($L=512$) Monte
Carlo steps were discarded (Table~\ref{Table1}). Measurements were taken every
five Monte Carlo steps. Between $15000$ ($L=64$) and $34692$ ($L=512$) measurements
were made for each random realization. These values were chosen in order to achieve
a good thermalization and then a sufficient sampling of the equilibrium probability
distribution. The number of measurements may appear at first sight unnecessarily large
since it may be argued that the average over randomness will smooth thermal
fluctuations. Smaller numbers of measurements have indeed led
to accurate values of the critical exponents in the past. However, in our
problem, preliminary tests have shown that a smaller number of Monte Carlo
steps leads to systematic deviations on the estimate of $\kappa$ due to the large
autocorrelation time of the interface.
\\

The measurements were then averaged over $N=5000$ different random configurations
of the exchange couplings $J_{ij}$ with the optimal ratio $r^*\simeq 4.0(3)$
determined by Jacobsen {\sl et al.}~\cite{Jacobsen09}. Since the random realizations
are uncorrelated, the measurements obtained for each of them are statistically
independent random variables so that the error can be estimated, according to
the central-limit theorem, as $\sqrt{\sigma^2/N}$ where $\sigma^2$ is the mean
square deviation of the observable among the random configurations.
Additional simulations have been performed for other values $r$ of the amplitude
of randomness. The same number of Monte Carlo steps were used but only
3000 random realizations were produced and the lattices sizes were limited
to $L=64$, 90 and 128.

\begin{table}[!ht]
\begin{center}
\begin{tabular}{@{}*{3}{l}}
$L$ & Thermalization & Measurement \\
\hline
64 & 25000 & 75000 \\
90 & 28750 & 86250 \\
128 & 33060 & 99185 \\
180 & 38015 & 114060 \\
256 & 43715 & 131165 \\
360 & 50270 & 150835 \\
512 & 57810 & 173460 \\
\hline
\end{tabular}
\end{center}
\caption{Number of Monte Carlo iterations used for thermalization
and measurements for the lattice sizes $L$ considered. For all lattice
sizes, the data were then averaged over 5000 random realizations.
The total number of Monte Carlo iterations is thus of order $10^9$
for $L=512$ and represents 3 years of single-CPU time (6 years for
all lattice sizes).}
\label{Table1}
\end{table}

\begin{center}
\begin{figure}
        \centerline{\psfig{figure=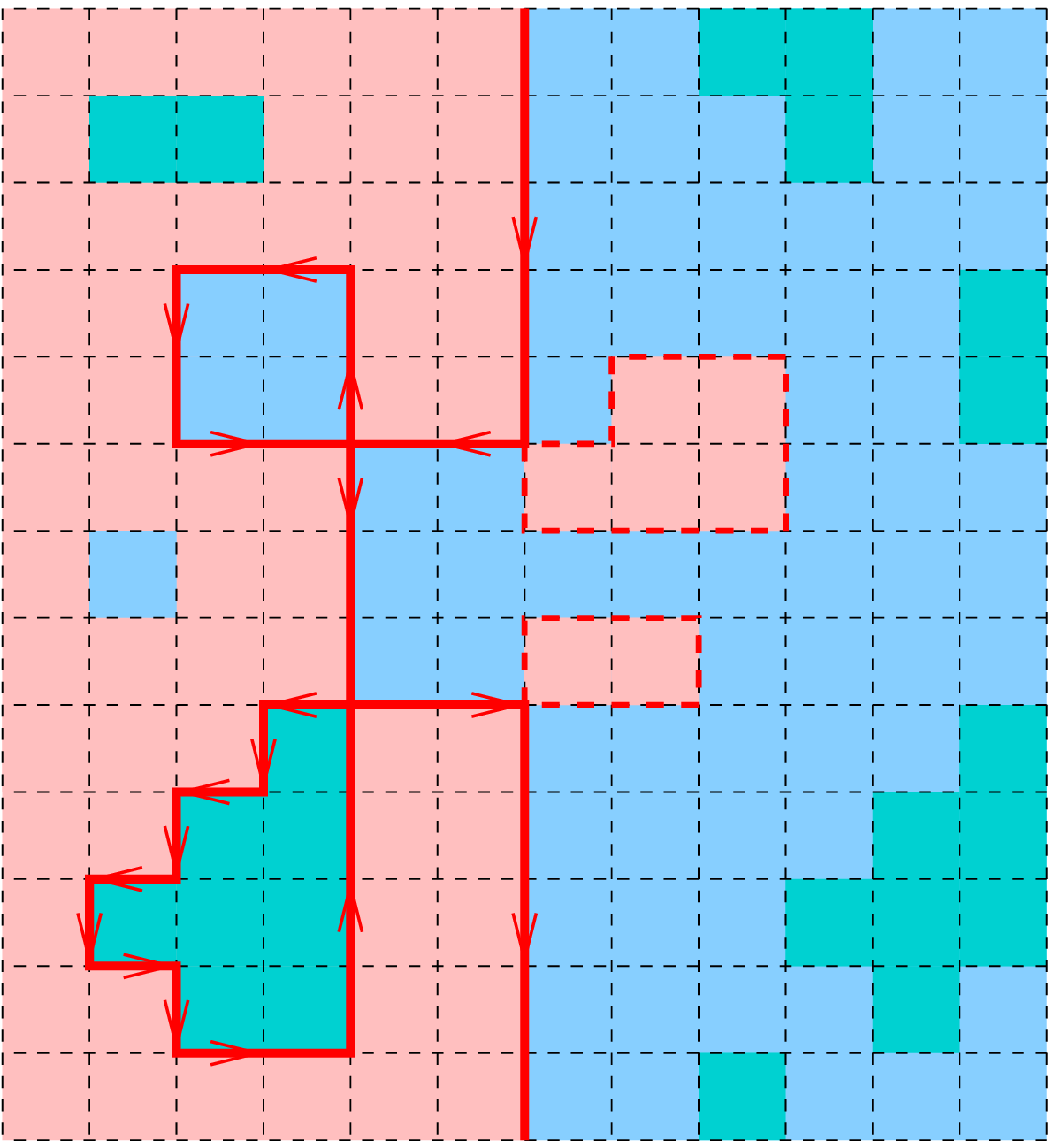,height=5cm}}
        \caption{Example of a spin configuration and the interface as determined
        by the algorithm. The three different spin states are depicted as
        pink, blue and green pixels. The interface is the bold red line.
        The dashed line corresponds to paths that are not followed by the
        algorithm.}
        \label{fig0}
\end{figure}
\end{center}

For a given spin configuration, the interface is determined by applying a
simple algorithm: starting from the upper point of the boundary where the
conditions change and thus the interface is fixed, the interface is built
on the dual lattice by going through the broken bonds between neighboring spins
(see figure \ref{fig0} for an example). The interface goes straight as long
as the spin at the right is in the same state as the left boundary and the one
at the left in one of the two other states. Ambiguous situations may arise: when
the interface cannot go straight, the spin at the right of the interface can be
in some cases in the same state as the left boundary whether the interface
turns on the right or on the left. The algorithm is thus not able to choose
one of the two possible directions~\footnote{To avoid these situations,
an hexagonal lattice, and thus a triangular dual lattice, is often considered.}.
On figure \ref{fig0}, four of these situations are depicted. We arbitrarily
imposed the rule that the interface should always turn on the right in
these cases ({\sl turn-right tie-breaking} rule).
As a consequence, the algorithm may fail to detect the correct interface.
Because of the rule of always turning on the right in case of ambiguity,
the interface may touch itself and thus form loops enclosing a cluster in a
different state. As shown on figure \ref{fig0}, this can only happen on the
left of the interface. Direct visualization of spin configurations shows that
the number of these loops is of order ${\cal O}(1)$ and that the cluster sizes
are most of the time limited to a few spins and are always much smaller than the
lattice size. 
\\

In order to gauge the influence of these unwanted loops on the scaling dimension
of the interface and on the left-passage probability, we estimated $\kappa$ for
the pure 3-state model and compared with the expected value $10/3$. As will be
seen in the following, the scaling dimension is not affected by the presence of
these loops. We note that the absence of noticeable effect on the scaling
dimension has also been reported recently in the case of the pure Ising model~\cite{Saberi10}.
It is not the case however for the left-passage probability. A small but
systematic deviation from the Schramm formula is observed on the left part
of the lattice. This deviation affects the estimation of $\kappa$. Following the
procedure presented in section \ref{sec3}, we obtain $\kappa\simeq 3.284$ for
$L=512$. Since the loops appears only on the left of the interface, it creates
a left-right asymmetry. Fortunately, the restriction of the interpolation of the data
with the Schramm formula to the right third of the lattice, i.e. for $(x,y)\in
[2L/3;L-1]\times [0;L-1]$ only turns out to give $\kappa\simeq 3.312$
for $L=512$ in much better agreement with the expected value $10/3$
for the pure model.

\section{Scaling of the interface length}
\label{sec2}
\begin{center}
\begin{figure}
        \centerline{\psfig{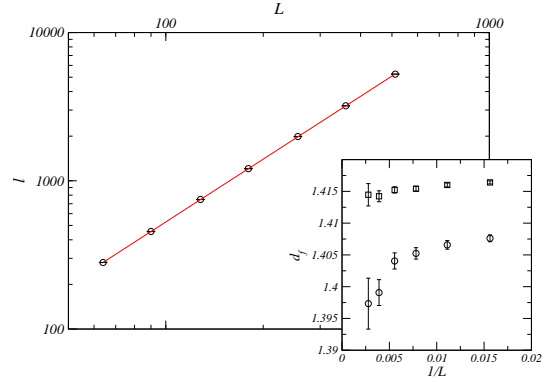}}
        \caption{Scaling of the interface length with respect to the
	lattice size $L$. The red line is the power-law fit taking into
	account all the points. In the inset, an effective fractal
	dimension $d_f$ obtained from a power-law fit of the points
	in the interval of lattice sizes $[L;512]$ is plotted with
	respect to $1/L$. The squares above correspond to the
	pure system and the circles to the random system with $r=4$.}
        \label{fig3}
\end{figure}
\end{center}

The scaling of the interface length was already determined by Jacobsen {\sl et al.}.
Besides reproducing their result, we present here a study of the influence of
disorder. As already mentioned, we will restrict ourselves to the interfaces
between spin clusters. For the pure model, we obtain $d_f\simeq 1.4164(4)$
corresponding to $\kappa\simeq 3.331(3)$, in very good agreement with the
expected value $10/3$. Our data for the disorder amplitude $r^*=4$, i.e. at the
random fixed point, is plotted on figure~\ref{fig3}. A simple power-law fit
$\ell\sim L^{d_f}$ leads to the fractal dimension $d_f\simeq 1.407(5)$ and
thus to the value $\kappa=3.26(4)$. However, a deviation from a pure power-law
can be observed, for example, by removing the smallest lattices sizes. In the
inset of figure \ref{fig3}, the fractal dimension obtained when taking into
account in the fit only the data with lattice sizes larger than $L$ is plotted
with respect to $1/L$. While the effective fractal dimension for the pure
system is relatively stable, a tendency to lower values of $d_f$ is clearly
observed in the random case. The last points are very noisy because the
interpolation is made only with the two or three largest lattice sizes.
The extrapolation to the thermodynamic limit is then delicate. Should the
two last points be interpreted as a real trend or a statistical fluctuation~?
Our final estimate will be $d_f\simeq 1.397(7)$ which correspond to
$\kappa\simeq 3.18(6)$. This value is less accurate but compatible within
error bars with the estimate given by Jacobsen {\sl et al.} ($3.208(24)$).

\begin{center}
\begin{figure}
        \centerline{\psfig{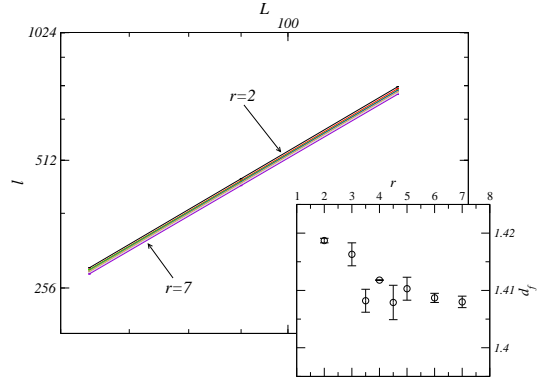}}
        \caption{Scaling of the length of the interface with respect to the
	lattice size $L$ for different disorder amplitudes $r$. In the inset,
	the fractal dimension $d_f$ is plotted with respect to $r$.}
        \label{fig2}
\end{figure}
\end{center}

The fractal dimension $d_f$ and thus $\kappa$ are sensitive to the
estimate $r^*$ of the disorder amplitude at the random fixed point.
As shown on figure \ref{fig2}, $d_f$ and thus $\kappa$ diminishes when
the disorder gets stronger. A linear interpolation of $d_f$ versus $r$ allows
for a rough estimation of the additional error $\Delta d_f$ due to the
uncertainty $\Delta r^*$ on $r^*$. With the value $\Delta r^*\simeq 0.3$
given by Jacobsen {\sl et al.}, we obtained $\Delta d_f\simeq 0.0006$
and thus $\Delta\kappa\simeq 0.005$, an order of magnitude smaller than
the statistical error.

\begin{center}
\begin{figure}
        \centerline{\psfig{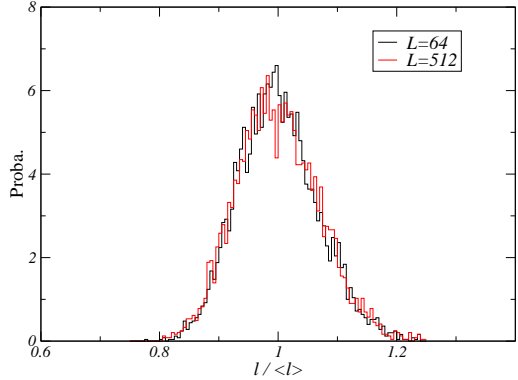}}
        \caption{Density probability of the length $l$ of the interface normalized
	by its average value $\langle l\rangle$. The two different curves correspond
	to different lattice sizes: $L=64$ (black) and $L=512$ (red).}
        \label{fig1}
\end{figure}
\end{center}

Besides the necessity to average over random realizations, numerical simulations
of random systems present the difficulty that an observable $\phi$ may not be
self-averaging. In such a situation, the average of $\phi$ over $N$
disorder realizations goes to $\langle\phi\rangle$ in the
limit $N\rightarrow +\infty$ but the relative fluctuations $R_\phi=[\langle\phi^2\rangle
-\langle\phi\rangle^2]/\langle\phi^2\rangle$ do not vanish. The probability
distribution of the length $\ell$ of the interface is shown on Figure~\ref{fig1}
for different lattice sizes $L$. When plotted with respect to $\ell/\langle\ell\rangle$,
the probability distribution appears the same for all lattice sizes. It means
that the mean square deviation $\langle\ell^2\rangle-\langle\ell\rangle^2$
scales as $\langle\ell\rangle^2$. The cumulant $R_\ell$ is thus constant: the interface
length is non self-averaging. 
Another important feature of random systems is that the typical event may differ
from the average. As seen on figure \ref{fig1}, the probability distribution
of the interface length is not symmetric and thus these events indeed differ
slightly. However, the distribution does not present any long tail with rare
events difficult to observe in numerical simulations. We rather observe a
distribution close to a Gaussian. Finally, random systems have the peculiar
property to possibly display multifractality: different moments of an observable
$\phi$ may present a different scaling behavior, i.e. $\langle \phi^n\rangle^{1/n}
\sim L^{x_\phi(n)}$ where $x_\phi$ depends of the order $n$ of the moment.
The first moments $\langle \ell^n\rangle^{1/n}$ have been computed. All fractal
dimensions $d_f(n)$ corresponding to the scaling behavior $\langle \ell^n
\rangle^{1/n}\sim L^{d_f(n)}$ are compatible within error bars. We conclude that
the interface length is not multifractal.

\section{Left-passage probability}
\label{sec3}
\begin{center}
\begin{figure}
        \centerline{\psfig{figure=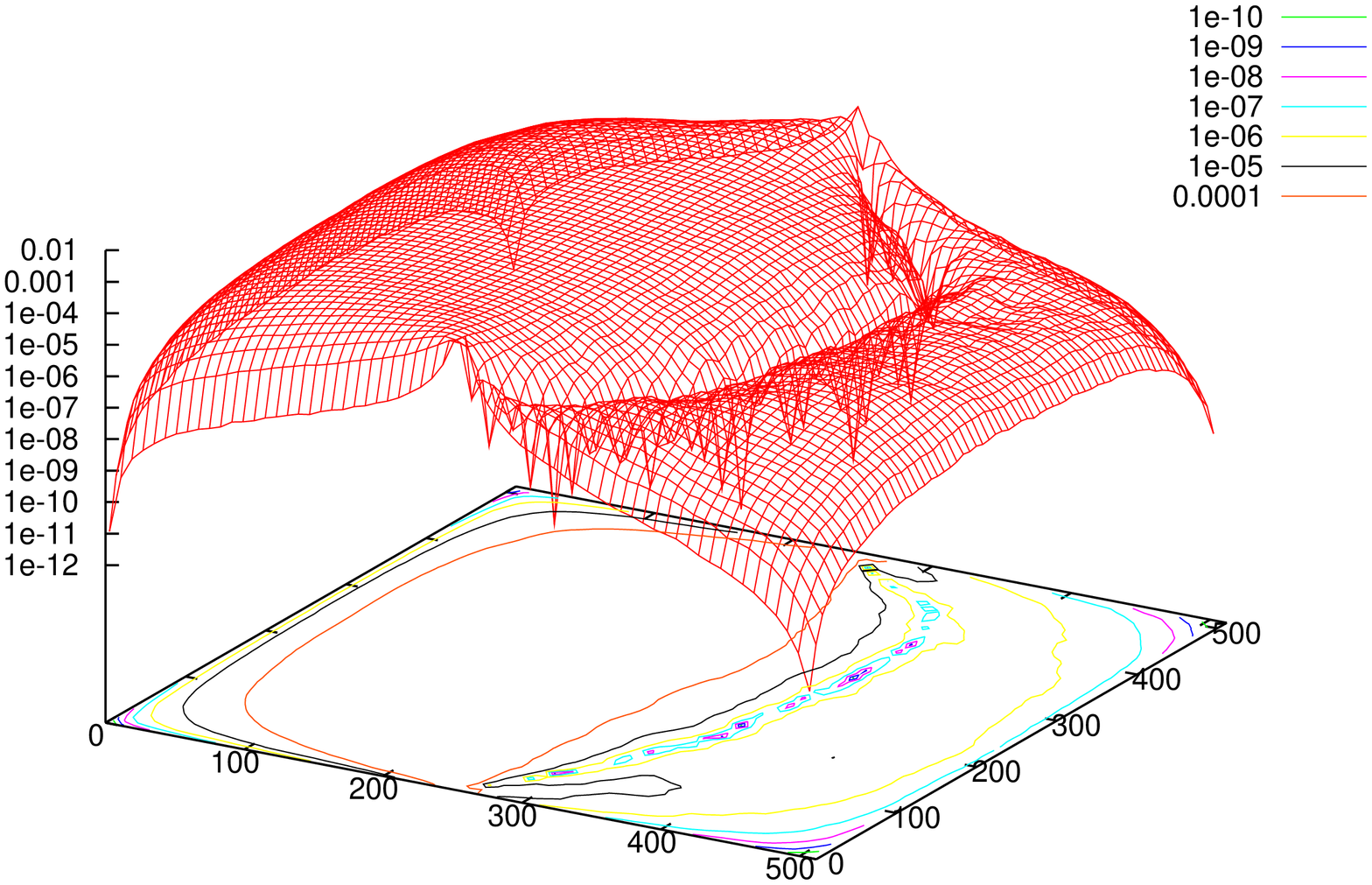,width=8.5cm,angle=-90}
	\psfig{figure=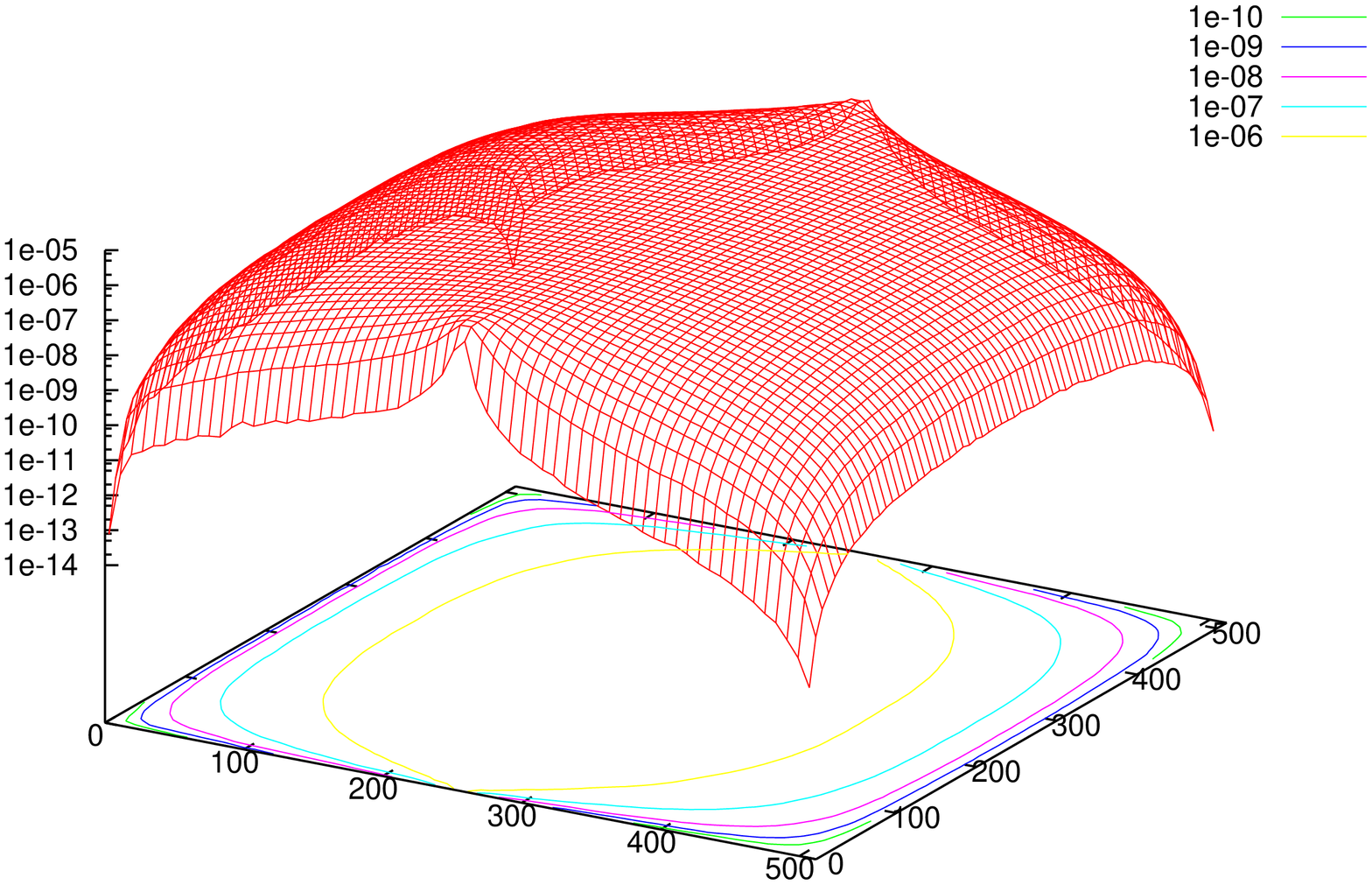,width=8.5cm,angle=-90}
        }\caption{On the right, square deviation $\chi^2(x,y)$ of the data
	with the Schramm formula with $\kappa=3.2452$ at each point of a
	$512\times 512$ lattice. The contour plot at the bottom shows that
	$\chi^2$ is of order $10^{-6}$ in the right part of the lattice. The
        largest discrepancy (of the order of $0.002$) is encountered close
        to the points where the interface meets the boundary.
	On the right, square of the statistical error $\Delta\wp^2(x,y)$ of the
	numerical estimate of the left-passage probability for the same system.
      }
        \label{fig4a}
\end{figure}
\end{center}

\begin{center}
\begin{figure}
        \centerline{\psfig{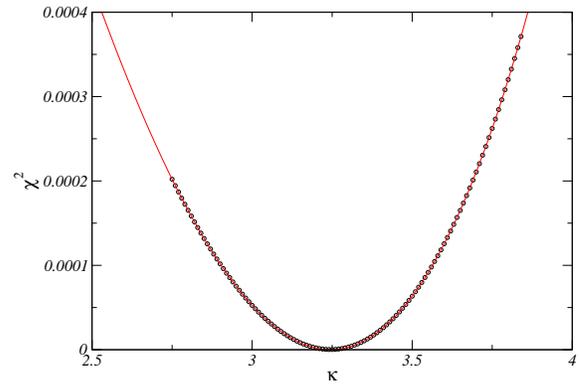}}
        \caption{Mean square deviation $\overline{\chi^2}$ of the data at the largest	
	lattice size $L=512$ with the Schramm formula with respect to $\kappa$. The line is
	the interpolation with a polynomial of fourth order ( $\overline{\chi^2} = 4.549281.10^{-3}
	-1.791033.10^{-3} \kappa -1.542747.10^{-5} \kappa^2 -1.270670.10^{-5} \kappa^3
	+ 1.677021.10^{-5} \kappa^4$) from which the minimum ($3.245225$) is then
	found numerically.}
        \label{fig4}
\end{figure}
\end{center}

We now determine $\kappa$ from the Schramm formula of the left-passage probability.
We follow the same procedure as \cite{Rieger09}. The left-passage probability $\wp(x,y)$
was estimated numerically for all points $(x,y)$ of the lattice. To extract $\kappa$,
the square deviation from the Schramm formula $S_\kappa(x,y)$ at any point of the
lattice
	$$\chi^2(x,y)=\big[\wp(x,y)-S_\kappa(x,y)\big]^2$$
was computed and then averaged only in the right third of the lattice:
	$$\overline{\chi^2}={1\over {2\over 3}L^2}\sum_{x=2L/3}^L
	\sum_{y=0}^L \chi^2(x,y).$$
The value of $\kappa$ leading to the smallest mean square deviation
is then searched as follows: the mean square deviation $\overline{\chi^2}$ is first
computed for a set of values $\kappa$ and then interpolated with a 4th-order polynomial
using a weight $1/\overline{\chi^2}$ in order to give a higher weight to the points
in the neighborhood of the minimum. The latter is then determined numerically by
searching for the root of the derivative of the 4th-order polynomial. An example of
this procedure is shown on figure \ref{fig4}. For the largest lattice size $L=512$,
the mean square deviation is minimal for $\kappa\simeq 3.2452$. The monitoring
of $\chi^2(x,y)$ shows that for this value of $\kappa$, the square deviation
is of order $10^{-6}$ in the right third of the lattice (figure \ref{fig4a}).
The square of the statistical error $\Delta\wp^2(x,y)$ is of the same order. 
The left-right asymmetry due to the loops is clearly seen: $\chi^2$ is larger
by two orders of magnitude in the left part of the lattice. The much better agreement
of the data with the Schramm formula in the right third of the lattice justifies
{\sl a-posteriori} the restriction of the interpolation to this region.

\begin{center}
\begin{figure}
        \centerline{\psfig{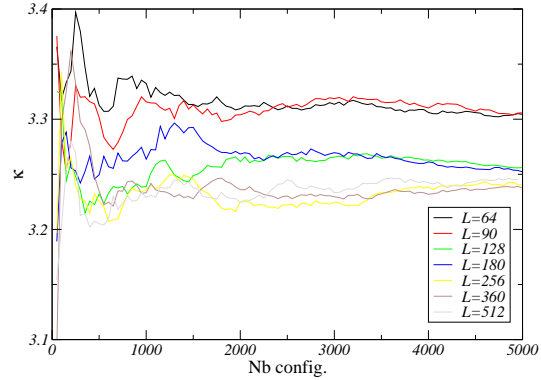}}
        \caption{Estimates of $\kappa$ when averaging over different
	numbers of random realizations.
	}\label{fig5}
\end{figure}
\end{center}

On figure \ref{fig5} (left), $\kappa$ is plotted versus the number of disorder realizations
taken into account in the average for different lattice sizes. No abrupt jump that would be
the signature of rare events with large contribution is observed. The estimates of $\kappa$
after $5000$ disorder realizations are compatible within statistical fluctuations
for lattice sizes larger than 128. Our final estimate is $\kappa\simeq 3.245(10)$.
For small lattice sizes, $\kappa$ takes a value close to 10/3, i.e. the one of the pure
3-state Potts model, suggesting that a cross-over between the pure and the random fixed
points occurs for $L^*\sim 100$.
The data have been analyzed using other fractions of the lattice. The estimate
of $\kappa$ tends to increase when going from a large fraction of the lattice to a
smaller one~: $3.222$ for an interpolation over the right half of the lattice, $3.255$
for an interpolation over the right quarter of the lattice. A compromise has to be
found between a too large fraction for which systematic deviations due to loops may be
important and a too small fraction where lattice effects due to the boundaries are
important. We note that, as expected, apllying blindly the interpolation procedure
to the whole lattice gives a significantly different estimate $\kappa\simeq 3.4$.

\begin{center}
\begin{figure}
        \centerline{\psfig{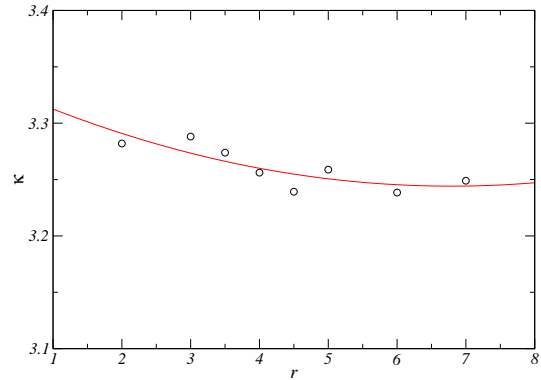}}
        \caption{Estimates of $\kappa$ for the lattice $L=128$ with respect to
        the disorder amplitude $r$. The red curve is a quadratic interpolation
	which gives $\kappa(r=1)=3.312$ for the pure model (we independently
	obtained $3.325$ for the pure model for this lattice size) and is
	intended to be only a guide to the eye. For small amplitude $r<r^*\simeq 4$,
	a cross-over with the pure value $10/3$ is observed. In the strong-disorder
	regime, $\kappa$ seems stable.	
	}
        \label{fig6}
\end{figure}
\end{center}

Like for the case of the interface length, we still need to show that a small
variation of the disorder amplitude will not induce a large variation of $\kappa$
so that $\kappa$ would still be compatible with the estimate $3.18(2)$
obtained from the scaling of the interface length. Simulations have been made
for other disorder amplitudes $r$ but only for lattice sizes $L\le 128$ and with 3000
disorder realizations. We have shown in the case of $r^*=4$ that this lattice size
still leads to a small overestimate of $\kappa$. Nevertheless, it should be possible to
get a correct order of magnitude of the variation of $\kappa$ with $r$. As shown
on figure \ref{fig6}, this variation remains small. With a linear interpolation, one
can estimate this variation to be $\Delta\kappa\sim 0.003$. 

\section*{Conclusions}
Although our interface presents loops due to the use of a simple right-turn tie-breaking
algorithm, we have shown that they do not affect the scaling dimension of the interface
in the pure case. However, systematic deviations between the left-passage probability
and the Schramm formula were observed. The parameter $\kappa$ can be nevertheless estimated
by restricting the interpolation to the right part of the lattice. 
In the case of the 3-state random bond Potts model, we have measured numerically the SLE
parameter $\kappa$  in two independent ways. First, we estimated $\kappa\simeq 3.18(6)$
from the fractal dimension $d_f$ of the interface. This value is compatible, although
less accurate, with the estimate obtained in~\cite{Jacobsen09}. Then, the left-passage
probability has been fitted with to the Schramm formula in the right third of the lattice.
We obtained $\kappa\simeq 3.245(10)$. These two compatible values provide further evidences
that chordal ${\rm SLE}_\kappa$ appropriately describes the geometric properties of interfaces
between spin clusters in the 3-state random Potts model. We have not observed the large
discrepancy observed in the case of the SOS model on a random substrate~\cite{Rieger09}.

\ack
The Institute Jean Lamour is Unit\'e Mixte de Recherche CNRS number 7198.
The author would like to express his gratitude to Sreedhar Dutta for having
patiently introduced him to the field of SLE.


\section*{References}
\begin{thebibliography}{64}
\bibitem{Cardy96} J. Cardy (1996)
{\sl Scaling and Renormalization in Statistical Physics}, Cambridge.

\bibitem{DiFrancesco}
P. Di Francesco, P, Mathieu, D, S\'en\'echal (1997),
{\sl Conformal Field Theory}, Springer.

\bibitem{Henkel99}
M. Henkel (1999) {\sl Conformal Invariance and Critical Phenomena}, Springer

\bibitem{Cardy05} J. Cardy (2005)
Annals of Phys. {\bf 318} 81,
{\tt cond-mat/0503313}.

\bibitem{Bauer06}
M. Bauer, D. Bernard (2006)
{\sl Phys. Rep.} {\bf 432} 115, {\tt math-ph/0602049}.

\bibitem{Lawler08} G. Lawler (2009)
{\sl Bull. Amer. Math. Soc.} {\bf 46} 35.

\bibitem{Potts}
R.B. Potts (1952) {\sl Proc. Camb. Phil. Soc.} {\bf 48} 106.

\bibitem{Smirnov01}  S. Smirnov (2001)
{\sl C.R. Acad. Sci. Paris} {\bf 333} 239,
{\tt arXiv:0909.4499}.

\bibitem{Smirnov06} S. Smirnov (2006)
{\sl Proc. Int. Congr. Math.} {\bf 2} 1421,
{\tt arXiv:0708.0032}

\bibitem{Ziff} D.A. Adams, L.M. Sander, R.M. Ziff (2010)
{\sl J. Stat. Mech.} P03004,
{\tt  arXiv:1001.0055}

\bibitem{Gamsa07} A. Gamsa, J. Cardy (2005)
{\sl J. Stat. Mech.} P12009,
{\tt math-ph/0509004}.

\bibitem{Gamsa05} A. Gamsa, J. Cardy (2007)
{\sl J. Stat. Mech.} P08020, 
{\tt arXiv:0705.1510}.

\bibitem{Gliozzi} F. Gliozzi, M.A. Rajabpour (2010)
{\sl J. Stat. Mech.} (2010) L05004,
{\tt arXiv:1003.3147}.

\bibitem{Berche}
C. Chatelain, B Berche (1998)
{\sl Phys. Rev. E} {\bf 58} R6899 
{\tt cond-mat/9810270}; C. Chatelain, B Berche
(1999) {\sl Phys. Rev. E} {\bf 60} 3853
{\tt cond-mat/9902212}.

\bibitem{Jacobsen09} J.L. Jacobsen, P. Le Doussal, M. Picco, R. Santachiara, K.J. Wiese
(2009) {\sl Phys. Rev. Lett. } {\bf 102} 070601,
{\tt arXiv:0809.3985}.

\bibitem{Bernard07} D. Bernard, P. Le Doussal, A. Middleton
(2007) {\sl Phys. Rev. B} {\bf 76} 020403(R),
{\tt cond-mat/0611433}.

\bibitem{Rieger09} K. Schwarz, A. Karrenbauer, G. Schehr, H. Rieger 
(2009) {\sl J. Stat. Mech.}, P08022,
{\tt arXiv:0905.4816}.

\bibitem{SWang}
R. Swendsen, J. Wang (1987) {\sl Phys. Rev. Lett.} {\bf 58} 86.

\bibitem{Hasenbusch91} M. Hasenbusch, S. Meyer (1991)
{\sl Phys. Rev. Lett.} {\bf 66}, 530

\bibitem{Saberi10} A.A. Saberi (2009) {\sl J. Stat. Mech.} P07030,
{\tt arXiv:0905.2451v2}


\end{thebibliography}
\end{document}